\documentclass[pre,reprint,twocolumn,superscriptaddress]{revtex4-2}

\pdfoutput=1

\usepackage{graphicx}
\usepackage[caption=false]{subfig}
\usepackage{amsmath}
\usepackage{amssymb}
\usepackage{color}

\newcommand{\am}[1]{{\textcolor{black}{#1}}}

\newcommand{\rev}[1]{{\textcolor{black}{#1}}}

\begin{document}
%%%

%%%%%General information

\begin{widetext}
\begin{center}{\textit{Published in Phys.~Rev.~E \textbf{106}, 064603 (2022). {\copyright}2022 American Physical Society.}}
\end{center}
\end{widetext}

\vspace{1.cm}

\title{Circular motion subject to external alignment under active driving ---\\ nonlinear dynamics and the circle map}

%%% Authors
\author{Andreas M. Menzel}
\email{a.menzel@ovgu.de}
\affiliation{Institut f{\"u}r Physik, 
Otto-von-Guericke-Universit{\"a}t Magdeburg, Universit{\"a}tsplatz 2, 39106 Magdeburg, Germany}

%%% Date
\date{\today}

\begin{abstract}
Hardly any real self-propelling or actively driven object is perfect. Thus, undisturbed motion will generally not follow straight lines but rather circular trajectories. We here address self-propelled or actively driven objects that move in discrete steps and additionally tempt to migrate towards a certain direction \am{by discrete angular adjustment}. \am{Overreaction in the angular alignment is possible.} This competition implies pronounced nonlinear dynamics including period doubling and chaotic behavior in a broad parameter regime. Such behavior directly affects the appearance of the trajectories, also during collective motion under spatial self-concentration. %Experimental observations should be straightforward. 
\end{abstract}

%\begin{document}

\maketitle
%\tableofcontents

\section{Introduction}

Most theoretical studies on self-propelled or actively driven objects consider idealized entities that migrate along straight lines when undisturbed \cite{vicsek1995novel, howse2007self, ten2011brownian, henkes2011active, ihle2013invasion, speck2014effective, elgeti2015physics, zottl2016emergent, romanczuk2016emergent, jin2021collective}. However, this idealization generally does not meet reality. During two-dimensional motion, for instance, on a substrate or near a surface, %persistent and invariable 
persistent imperfections of otherwise undisturbed objects can lead to %motion along 
circular trajectories. 
To study and emphasize the resulting effects, artificial so-called circle swimmers that show extreme deviations from axially symmetric shapes \cite{kummel2013circular, ten2014gravitaxis, bechinger2016active} or nonsymmetric 
%chiral 
vibrated hoppers \cite{kubo2015mode, deblais2018boundaries} were generated and investigated. Similarly, several biological objects show circle swimming, for example, alga cells that feature a defect in one of their usually two driving arms \cite{brokaw1982analysis, kamiya1984submicromolar}. \am{If motion is achieved by rotational effects, such as the rotation of flagella in \textit{Escherichia coli} bacteria, hydrodynamic coupling to nearby substrates and surfaces can lead to circular trajectories \cite{lauga2006swimming, lemelle2010counterclockwise, di2011swimming, lemelle2013curvature, hu2015physical, daddi2018hydrodynamic}.} Besides, spontaneous symmetry breaking can induce persistently bent trajectories \cite{ohta2009deformable, kruger2016curling}. 
\am{While the spontaneous emergence of curved paths of propagation %and the possible reversal of their sense 
is an important topic %and has been addressed in various contexts 
\cite{casiulis2022emergent}, we here concentrate on objects featuring an inherent, permanent tendency of circular motion. Thus spontaneously emerging chiral motion is only covered by our consideration as long as it is sufficiently persistent over the time of observation.}

The dynamics of such active objects propagating along circular trajectories have been analyzed theoretically in quite some detail \cite{van2008dynamics, tarama2012spinning, lowen2016chirality, hoell2017dynamical, liebchen2017collective, kurzthaler2017intermediate, liao2018clustering, caprini2019active, huang2020dynamical, lowen2020inertial, liao2021emergent}. \am{It is common to implement in a minimal approach the tendency of circular motion using a constant torque or angular frequency that affects angular orientation.} \rev{In the mentioned theoretical studies \cite{van2008dynamics, tarama2012spinning, lowen2016chirality, hoell2017dynamical, liebchen2017collective, kurzthaler2017intermediate, liao2018clustering, caprini2019active, huang2020dynamical, lowen2020inertial, liao2021emergent}, the dynamics was considered as continuous in time.}

\rev{Yet and thus, another} aspect is that many active objects move by successive discrete \am{steps or other} events of propulsion. Obviously, this applies to humans and animals performing discrete steps or jumps, but to some degree also %, for instance, 
to birds and fish that flap with their wings and fins \cite{couzin2002collective, buhl2006disorder, ballerini2008interaction, tunstrom2013collective}. On smaller scales, actively driven hoppers on vibrating plates move by discrete bounces \cite{blair2003vortices, deseigne2010collective, scholz2021surfactants}. To some extent, the run-and-tumble motion of microscopic self-driven swimming objects like the bacterium \textit{Escherichia coli} \cite{berg1972chemotaxis} can be interpreted as discrete stepwise motion between individual tumbling events\am{, as can the stop-shock-run dynamics of the marine alga \textit{Pyramimonas octopus} \cite{wan2018time}}. The famous theoretical Vicsek model was formulated for discrete steps of motion \cite{vicsek1995novel, gregoire2004onset}. \am{Variants of this model are still under intense investigation, concerning, for instance, the recent discovery of additional dynamic phases \cite{kursten2020dry} or collective behavior in the presence of obstacles \cite{codina2022small}.} It has been demonstrated that finite step size can lead to unexpected types of behavior such as unidirectional laning \rev{and migrating cluster crystals} \cite{menzel2013unidirectional}. \rev{These phenomena result from the overreaction that becomes possible from performing discrete and finite steps.}

\am{As a third ingredient}, %and in contrast to what is assumed in many theoretical models, 
self-propelled objects frequently do not just migrate without any goal. Often, they tend to head towards a certain direction, for instance, when microorganisms turn towards or away from light \cite{schaller1997chlamydomonas} or humans follow escape routes \cite{helbing2000simulating}. 

Here, we combine the above ingredients. That is, we analyze the dynamics of individual self-propelled or actively driven objects that move in discrete steps, would generally follow circle-like trajectories, if undisturbed, but simultaneously tend to migrate towards a certain direction, for instance, to reach a (remote) target. \rev{We are interested in the various modes of behavior that result from such a combination.} It turns out that \rev{this specific combination of the mentioned aspects of behavior as addressed in the following}, already for simple individual objects, leads to complex dynamics. Specifically, intermediate period doubling and chaotic motion emerges, depending on the strength of alignment. 

%In the following, we first introduce the equations of motion in Sec.~\ref{sec_model}. Next, we present resulting trajectories for different strengths of alignment in Sec.~\ref{sec_traj}. Afertwards, we turn to more quantitative measures concerning these trajectories and the motion in general in Sec.~\ref{sec_quant}. Some aspects of collective motion are addressed in Sec.~\ref{sec_coll}, before we conclude in Sec.~\ref{sec_concl}. 

\section{Equations of motion}
\label{sec_model}

\begin{figure}
\centerline{\includegraphics[width=.9\columnwidth]{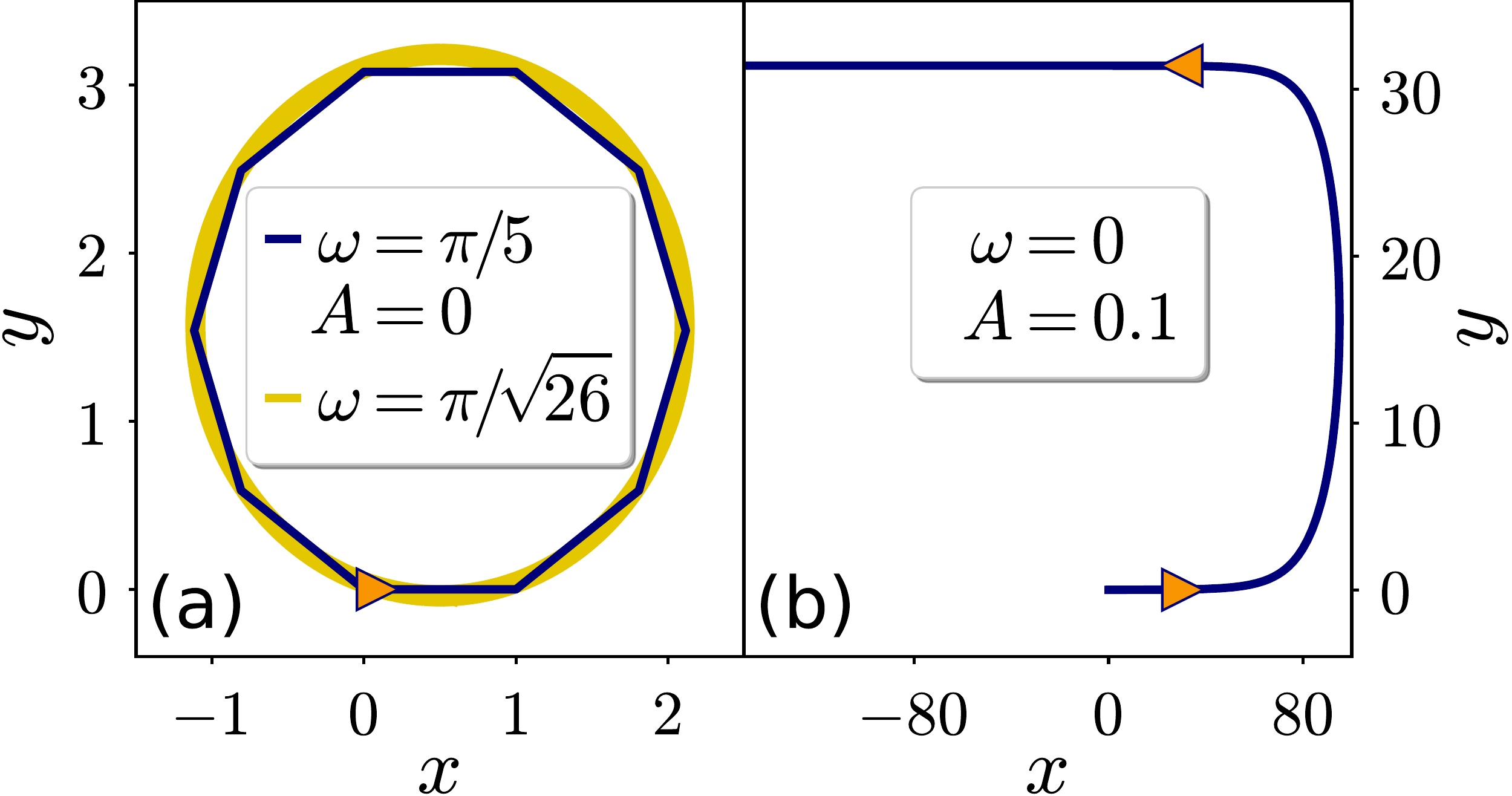}}
\caption{
Trajectories %of discrete types of motion 
for two extreme cases. % of Eqs.~(\ref{eq_varphi})--(\ref{eq_y}). 
(a) Vanishing tendency of alignment $A=0$ for angular frequency $\omega=\pi/5$ \am{(darker polygon-shaped line)}. We observe a closed regular trajectory of discrete steps. \am{Superimposing many cycles for $\omega=\pi/\sqrt{26}$, a circular shape of finite thickness emerges (brighter, thicker line in the background).} (b) Vanishing angular frequency $\omega=0$ for alignment tendency $A=0.1$. In our case, the driven object tends to orient its migration direction along $-\mathbf{\hat{x}}$, %aiming at an orientation 
corresponding to $\varphi=\pi$. Triangles indicate the direction of motion. 
}
\label{fig_simpletraj}
\end{figure}

We consider the \am{two-dimensional} motion of a self-propelled or actively driven object, for instance, when moving on a substrate. The present direction of motion is parameterized by an angle $\varphi$ that we measure from the $x$-axis of our Cartesian coordinate frame. %This angle 
$\varphi$ is updated from each time step $n$ to the next one %time step $n+1$ 
according to %the relation
\begin{equation}\label{eq_varphi}
\varphi_{n+1} = \varphi_n + \omega + A\,\sin\varphi_n. 
\end{equation}
Once $\varphi$ %as $2\pi$-periodic, that is, we 
leaves the interval $[0,2\pi[$, we
map it back by adding $\pm2\pi$. % once it leaves this interval. 
\am{In this description,} $\omega$ determines the angular frequency by which $\varphi$ changes in a discrete way, scaled by the duration of the time step. \am{This is consistent with setting a constant torque in previous continuous models of circular self-propulsion \cite{van2008dynamics, tarama2012spinning, lowen2016chirality, liebchen2017collective, kurzthaler2017intermediate, liao2018clustering, caprini2019active, huang2020dynamical, lowen2020inertial, liao2021emergent}.} \am{To illustrate our results in the following}, we %here %decided to 
\am{mostly select one specific value of} %it to 
$\omega=\pi/5$, %throughout. This leads 
leading to closed kinked trajectories %that feature kinks and 
of the shape of a regular polygon, see Fig.~\ref{fig_simpletraj}(a), in \am{analogy to the} smooth circles \am{appearing} in the continuous case. 
\am{In fact, polygon-shaped trajectories have recently been triggered for certain light-sensitive microswimmers \cite{tsang2018polygonal}.}
Testing other commensurate values of $\omega$, we generally recovered the qualitative signature of our results. 
\am{When incommensurate values of $\omega$ are considered, particularly irrational multiples of $\pi$, the regular polygon in Fig.~\ref{fig_simpletraj}(a) appears smeared to a circle of finite thickness when the trajectory over many cycles overlays itself. We illustrate an example of $\omega=\pi/\sqrt{26}$ by the brighter curve in the background of Fig.~\ref{fig_simpletraj}(a). In the following cases, the curves for the two values of $\omega$ appear qualitatively similar, apart from quantitative deviations due to the slightly shifted magnitude of $\omega$ from $\pi/5$ to $\pi/\sqrt{26}$.} \rev{Depending on the magnitude of the tendency of alignment $A$, these quantitative deviations can, however, become quite significant, as illustrated below.} 

\am{Most importantly,} the parameter $A$ heading the last term in Eq.~(\ref{eq_varphi}) \am{is associated with the} strength of external alignment tendency. For straight-moving objects, that is, for $\omega=0$, $A>0$ \am{in the continuous case will generally induce} a heading towards the negative $x$-direction $-\mathbf{\hat{x}}$ given by $\varphi=\pi$. \am{Such an alignment could be induced, for example, magnetically for magnetic self-propelled Janus particles \cite{baraban2013control, baraban2013fuel, demirors2018active} or magnetotactic bacteria \cite{blakemore1975magnetotactic, blakemore1982magnetotactic, klumpp2016magnetotactic}. Another situation corresponds to a remote target far away in the direction $-\mathbf{\hat{x}}$ that the self-propelled or actively driven object tries to reach. This could be a source of nutrient or the only exit under confinement. In our discretized consideration, this tendency of alignment along the direction $-\mathbf{\hat{x}}$ becomes apparent for $\omega=0$ at small values of $A$}, see Fig.~\ref{fig_simpletraj}(b). \am{For larger magnitudes of $A$, overreactions in alignment towards $-\mathbf{\hat{x}}$ can occur in our discretized description. %, if the initial alignment becomes too strong. 
Qualitative effects on the dynamics of self-propelled particles caused by such overreactions can be substantial, as outlined before in a different context \cite{menzel2013unidirectional}.}
\am{We remark that, in the continuous case, variants of Eq.~(\ref{eq_varphi}) have been studied, for instance, in the context of flagellar synchronization \cite{goldstein2009noise} and of the transition to upstream swimming of sperm cells \cite{tung2015emergence} and bacteria \cite{mathijssen2019oscillatory} near surfaces.}

The actual motion of our object %then 
follows from a propulsive step 
\begin{eqnarray}
x_{n+1} &=& x_n + \cos\varphi_n, 
\label{eq_x}
\\
y_{n+1} &=& y_n + \sin\varphi_n.
\label{eq_y}
\end{eqnarray}
In these relations, we have scaled the spatial positioning by the %product of duration 
duration of the time step and the migration speed. Both are assumed to be constant \cite{vicsek1995novel}. 

We solve Eqs.~(\ref{eq_varphi})--(\ref{eq_y}) %of motion 
by direct numerical iteration. As an initial condition, we start from a heading towards %the positive $x$-direction 
$+\mathbf{\hat{x}}$ %, that is, from 
($\varphi_0=0$), % in all cases, 
unless mentioned otherwise. 
\am{We note that Eq.~(\ref{eq_varphi}) basically represents an overdamped type of dynamics for the angular variable, which sets the velocity orientation. Thus, this approach cannot reproduce phenomena that originate when inertial effects associated with angular momentum become important, as described, for instance, in the context of the turning dynamics of flocks of birds \cite{attanasi2014information, cavagna2015flocking} or so-called ``microflyers'' \cite{lowen2020inertial, menzel2022statistics}.}

%\am{Of course, real systems are not perfectly deterministic and fluctuations in the system parameters will emerge over time. Frequently, these variations are modeled by adding terms of stochastic noise to the equations of motion, particularly to the angular iteration %in Eq.~(\ref{eq_varphi}) 
%\cite{vicsek1995novel}. Here, on purpose, we do not include such contributions. Our goal is to outline the nonlinear and chaotic dynamics that already emerge from the combined equations of motion as listed above, without the need for adding stochastic contributions.}

\section{Different types of trajectories}
\label{sec_traj}

\begin{figure}
\centerline{\includegraphics[width=.9\columnwidth]{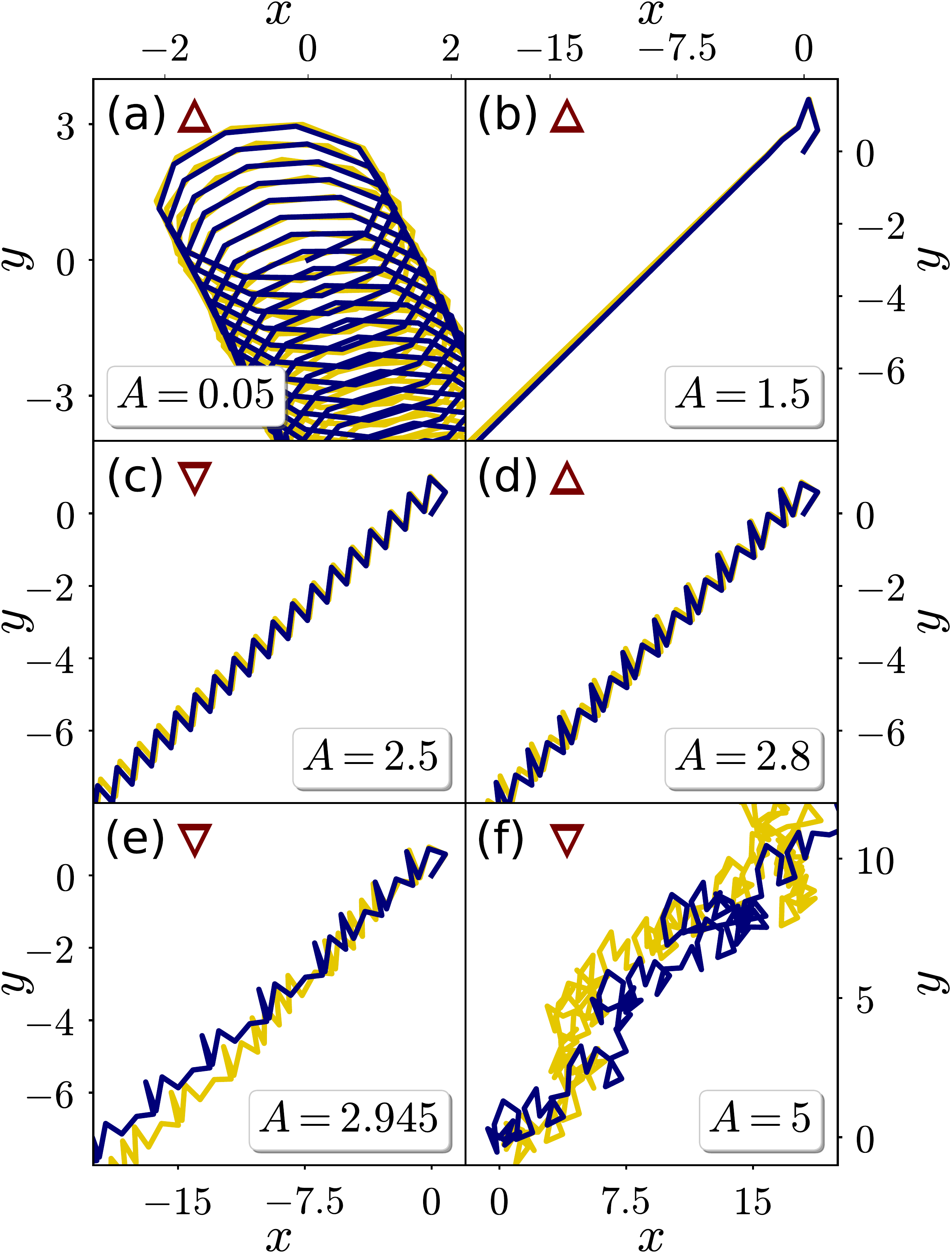}}
\caption{
Trajectories according to Eqs.~(\ref{eq_varphi})--(\ref{eq_y}) for combinations of nonvanishing angular frequency $\omega=\pi/5$ and tendency of alignment $A\neq0$ \am{(darker lines)}. (a) Cycloidal-like trajectories are observed for small but nonvanishing $A$, here $A=0.05$. (b) For $A=1.5$ we find straight motion with a constant angle of deviation from $-\mathbf{\hat{x}}$. (c) From there, for increasing $A$, regular zigzag-like trajectories appear, as depicted for $A=2.5$. We observe non-binary zigs and zags with further increasing $A$, here for (d) $A=2.8$ and (e) $A=2.945$. (f) Chaotic motion emerges, for example, for $A=5$. The triangular arrows \am{next to each panel label} indicate whether the top or bottom abscissa scale applies. \am{Brighter lines in the background show, for comparison, results for $\omega=\pi/\sqrt{26}$.} 
}
\label{fig_complextraj}
\end{figure}

When evaluating Eqs.~(\ref{eq_varphi})--(\ref{eq_y}), we made an unanticipated observation. We expected on the basis of Figs.~\ref{fig_simpletraj}(a) and (b) a combination of cyclic motion with superimposed drift, that is, a discrete version of a cycloidal-like trajectory. In fact, such types of trajectories are found for various parameter combinations, see Fig.~\ref{fig_complextraj}(a). % for an example. 
Yet, further types of trajectory result, such as straight lines, although oblique to the direction of preferred alignment, see Fig.~\ref{fig_complextraj}(b). Moreover, regular zigzag-like trajectories appear, see Fig.~\ref{fig_complextraj}(c), besides trajectories of more complex %than simple binary 
zigs and zags, see Figs.~\ref{fig_complextraj}(d) and (e). 
In many parameter regimes, we found very irregular types of trajectory, see Fig.~\ref{fig_complextraj}(f), although we are working with a very simple deterministic system. 

\begin{figure}
\centerline{\includegraphics[width=.9\columnwidth]{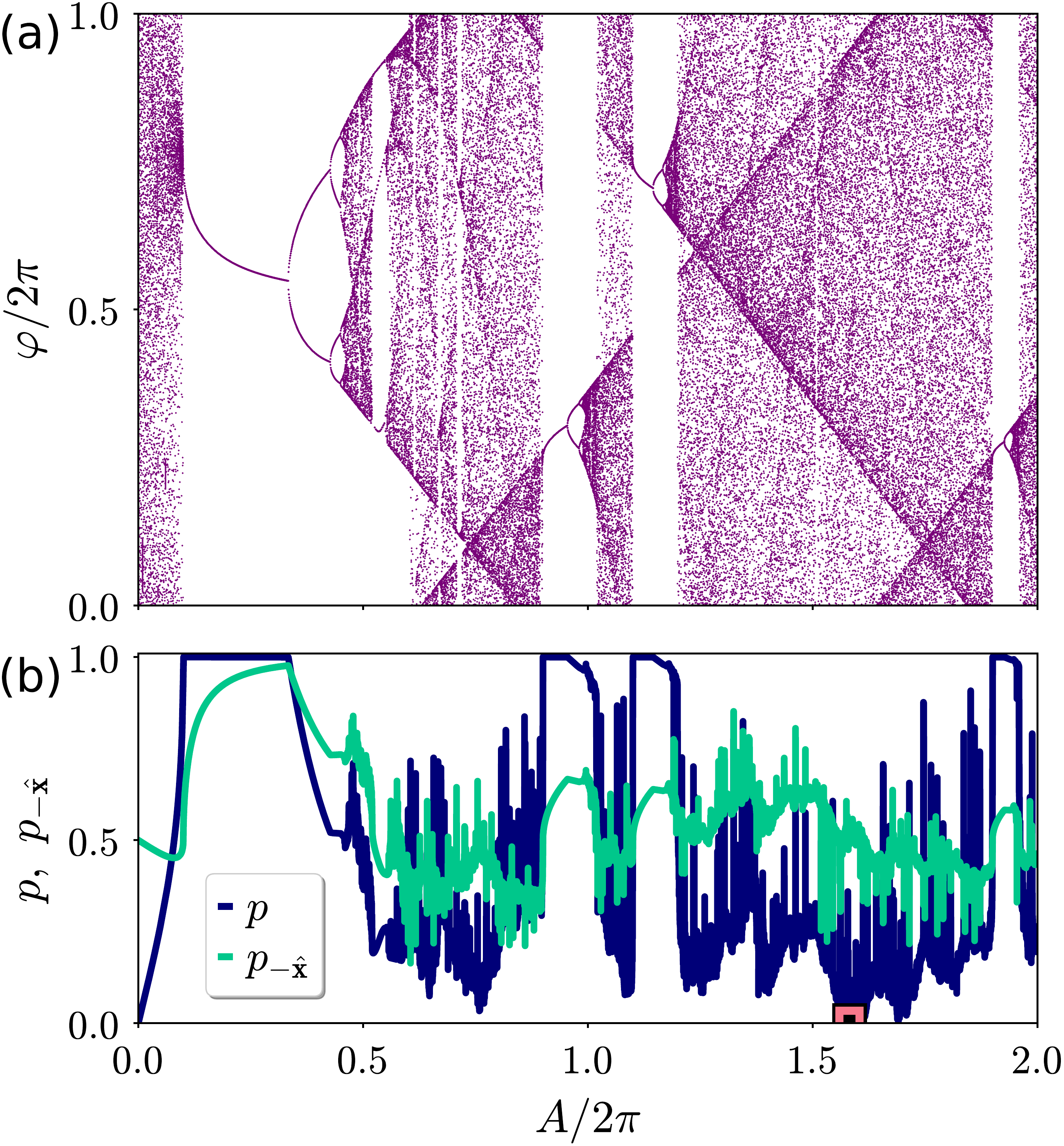}}
\caption{
Overview of the circle map. (a) Numerical evaluation of Eq.~(\ref{eq_varphi}) for $\omega=\pi/5$ in the interval $0\leq A\leq4\pi$. While periodic behavior exists at $A=0$ and in several additional intervals of $A$, we observe period doubling and pronounced intervals of chaotic motion as well. This map was obtained by plotting the result of $10^4$ iterations of Eq.~(\ref{eq_varphi}) for each value of $A$, after previous $10^4$ steps of iteration. (b) Order parameters $p$ and $p_{\am{-\mathbf{\hat{x}}}}$ according to Eqs.~(\ref{eq_pd}) and (\ref{eq_prd}) %, respectively, 
for $N_{\mathrm{step}}=10^4$. We mark by the cut square the relatively low value of the order parameter $p\approx0.0066$ at $A=9.940441$ used %to obtain the results 
in Figs.~\ref{fig_exampledrift} and \ref{fig_collective}. 
}
\label{fig_arnold}
\end{figure}

How can we understand these observations? 
In fact, it turns out that Eq.~(\ref{eq_varphi}) corresponds to a famous minimal model studied in the context of nonlinear dynamics. It is the so-called Arnol'd circle map \cite{arnold2009circle}. \am{Aspects of its properties, for example the resulting bifurcations and transitions to chaotic behavior, have been analyzed in detail \cite{glass1982fine, glass1983bifurcation, ostlund1983universal, jensen1984transition, boyland1986bifurcations}.} %, although generally addressed after rescaling the angles and parameters by a factor of $2\pi$. % and investigated by Arnol'd with $\cos$ instead of $\sin$. 
Here, \am{the equation, instead of being formulated as a purely mathematical model,} arises naturally %and directly 
from \am{a physical context when combining} circular motion and the tendency \am{of moving} towards a preferred direction. 
While the circle map does feature periodic intervals, phenomena of period doubling and chaotic behavior emerge as well. We include an evaluation of Eq.~(\ref{eq_varphi}) for the interval $0\leq A\leq4\pi$ in Fig.~\ref{fig_arnold}(a) \am{for illustration}. 
\am{In our context, these results get transformed into trajectories in the two-dimensional plane as described in the following.}

At precisely $A=0$, periodic behavior is found, which leads to the regular polygon-like trajectory in Fig.~\ref{fig_simpletraj}(a) for $\omega=\pi/5$. For slightly larger values %When $A$ is increased from 
$A\gtrsim0$, the distance between the cyclic values of $\varphi$ starts to %slightly 
deviate 
from exactly $\omega=\pi/5$. Consequently, the spectrum of $\varphi$-values in Fig.~\ref{fig_arnold}(a) gets broadened, corresponding to cycloidal-like trajectories as in Fig.~\ref{fig_complextraj}(a). 

The map in Fig.~\ref{fig_arnold}(a) illustratively explains the remaining behavior observed in Fig.~\ref{fig_complextraj}. 
Intermediate $A$-intervals are characterized by a single recurrent value of the angle $\varphi$. % in Fig.~\ref{fig_arnold}. 
These intervals indicate a fixed angular orientation, yet generally oblique to the preferred direction $-\mathbf{\hat{x}}$, see Fig.~\ref{fig_complextraj}(b). Apparently, the tendencies of rotation set by $\omega$ and of alignment set by $A$ %in this case 
here balance each other, %leading as a compromise 
compromising to an offset angle %of 
$\varphi\neq\pi$. Such intervals are with increasing $A$ followed in Fig.~\ref{fig_arnold}(a) by events of period doubling. After the first %event of period 
doubling, two alternating angles are associated with zigzag-type motion, see Fig.~\ref{fig_complextraj}(c). More complex trajectories result after further period doubling, which translates into different zigs and zags, see Figs.~\ref{fig_complextraj}(d) and (e). 
Finally, the broad intervals of obviously chaotic behavior in Fig.~\ref{fig_arnold}(a) are reflected by the rather irregular appearance of associated trajectories, see Fig.~\ref{fig_complextraj}(f).

\begin{figure}
\centerline{\includegraphics[width=1.\columnwidth]{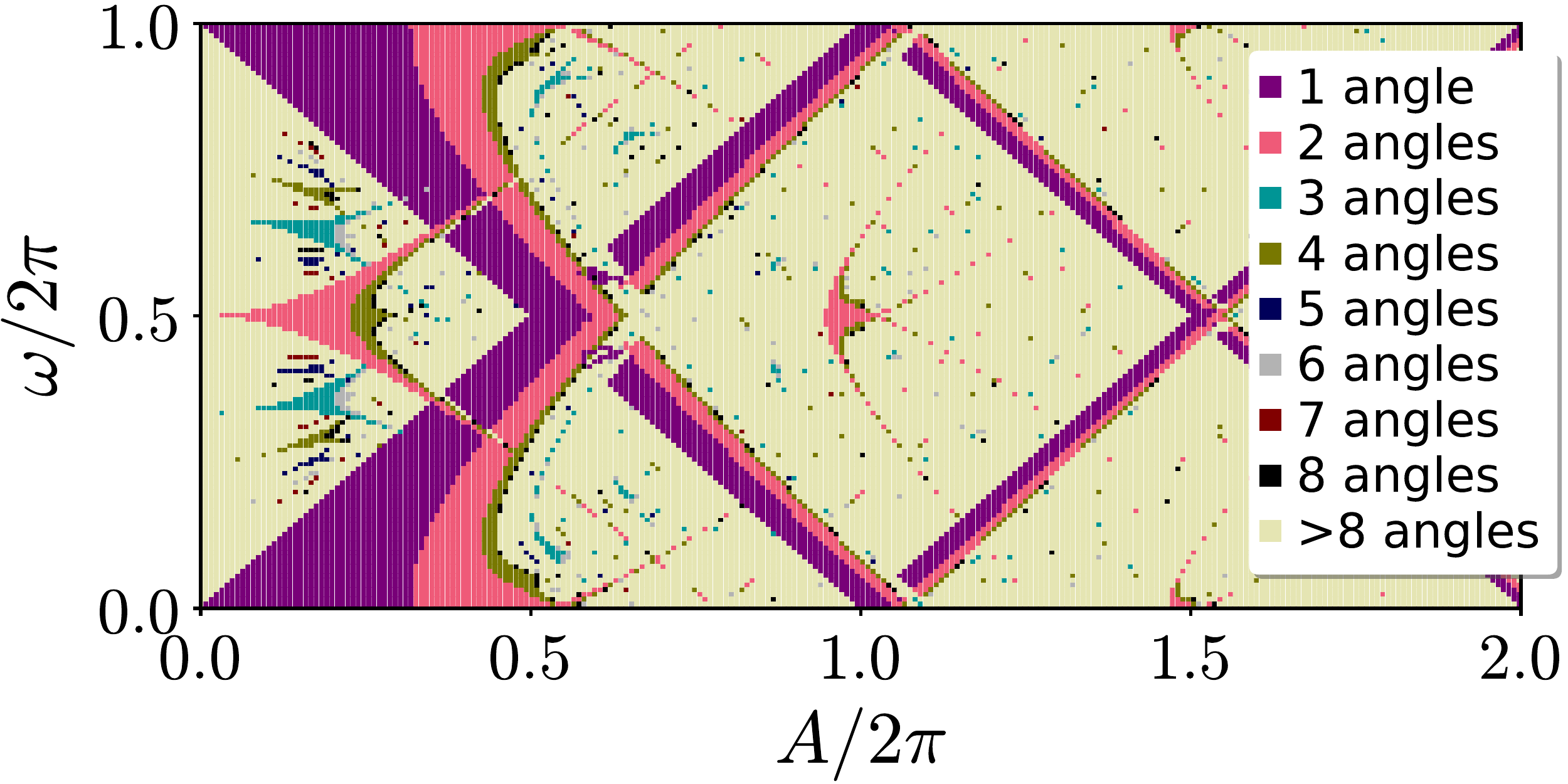}}
\caption{
\rev{Overview on the different dynamic regimes in the steady state observed when varying the angular frequency $\omega$ and the tendency of alignment $A$. We distinguish between types of behavior for which repeated evaluations of Eq.~(\ref{eq_varphi}) lead to one [see Fig.~\ref{fig_complextraj}(b)], two [see Fig.~\ref{fig_complextraj}(c)], three, four [see Fig.~\ref{fig_complextraj}(d)], five, six, seven, eight [see Fig.~\ref{fig_complextraj}(e)], or more different values of $\varphi$. The latter case is predominantly associated with chaotic behavior [see Fig.~\ref{fig_complextraj}(f)] in our resolution. We set the tolerance when testing whether angles are identical to $\pm\pi/1000$. 
}
}
\label{fig_phasediag}
\end{figure}

\rev{Mainly, we focus on the values of the angular frequency $\omega$ as indicated above and in the captions of Figs.~\ref{fig_simpletraj}--\ref{fig_arnold}. Yet, to briefly demonstrate how the different dynamic regimes as illustrated in Figs.~\ref{fig_complextraj} and \ref{fig_arnold} are affected when varying $\omega$, we include Fig.~\ref{fig_phasediag}. There, we test for the dynamic behavior for which Eq.~(\ref{eq_varphi}) in the steady state evaluates to one, two, three, four, five, six, seven, eight, or more different values of $\varphi$ upon iteration. To our resolution, the latter case may predominantly be ascribed to chaotic behavior.}

\section{Characterizing the trajectories and types of motion}
\label{sec_quant}

%Generally, we are used to quantifying irregular motion of self-propelled objects by their mean-squared displacements and diffusion coefficients \cite{howse2007self, ten2011brownian}. In our case, however, the motion described by Eqs.~(\ref{eq_varphi})--(\ref{eq_y}) is completely deterministic. Thus ensemble averages over different stochastic realizations are not meaningful. 

As a first step of \am{quantifying the observed types of motion}, we define an order parameter 
\begin{equation}\label{eq_pd}
p=\frac{1}{N_{\mathrm{step}}}\left[\left(\sum_{n=1}^{N_{\mathrm{step}}}\cos\varphi_n\right)^{\!\!2} + \left(\sum_{n=1}^{N_{\mathrm{step}}}\sin\varphi_n\right)^{\!\!2} \right]^{1/2},
\end{equation}
where the sums run over $N_{\mathrm{step}}$ subsequent time steps. This parameter vanishes, $p=0$, in the absence of any persistent 
net drift towards a preferred direction. Conversely, $p=1$ signals persistent and fully directed motion towards a certain direction. % without any intermediate deviation. 
Therefore, the drift parameter $p$ measures how effectively the object moves forward. % and can be interpreted as a drift parameter. 

In Fig.~\ref{fig_arnold}(b), $p$ is plotted as the darker line. It starts from $p=0$ at $A=0$, in agreement with the closed trajectory that does not imply any average displacement. With increasing tendency of alignment, the object starts to drift towards a certain direction and $p$ increases with increasing $A$. 
The motion is most effective and fully directed along one axis in intervals of only one angle in Fig.~\ref{fig_arnold}(a), %. In these intervals 
where $p=1$. Period doubling reduces $p$, while chaotic motion can push it to very low values. 
%
%Although $p$ may become very small during chaotic motion, 
Still, we always found $p>0$ in these intervals\am{, implying that} a net drift motion is associated with the irregularly shaped trajectories. 

\begin{figure}
\centerline{\includegraphics[width=.9\columnwidth]{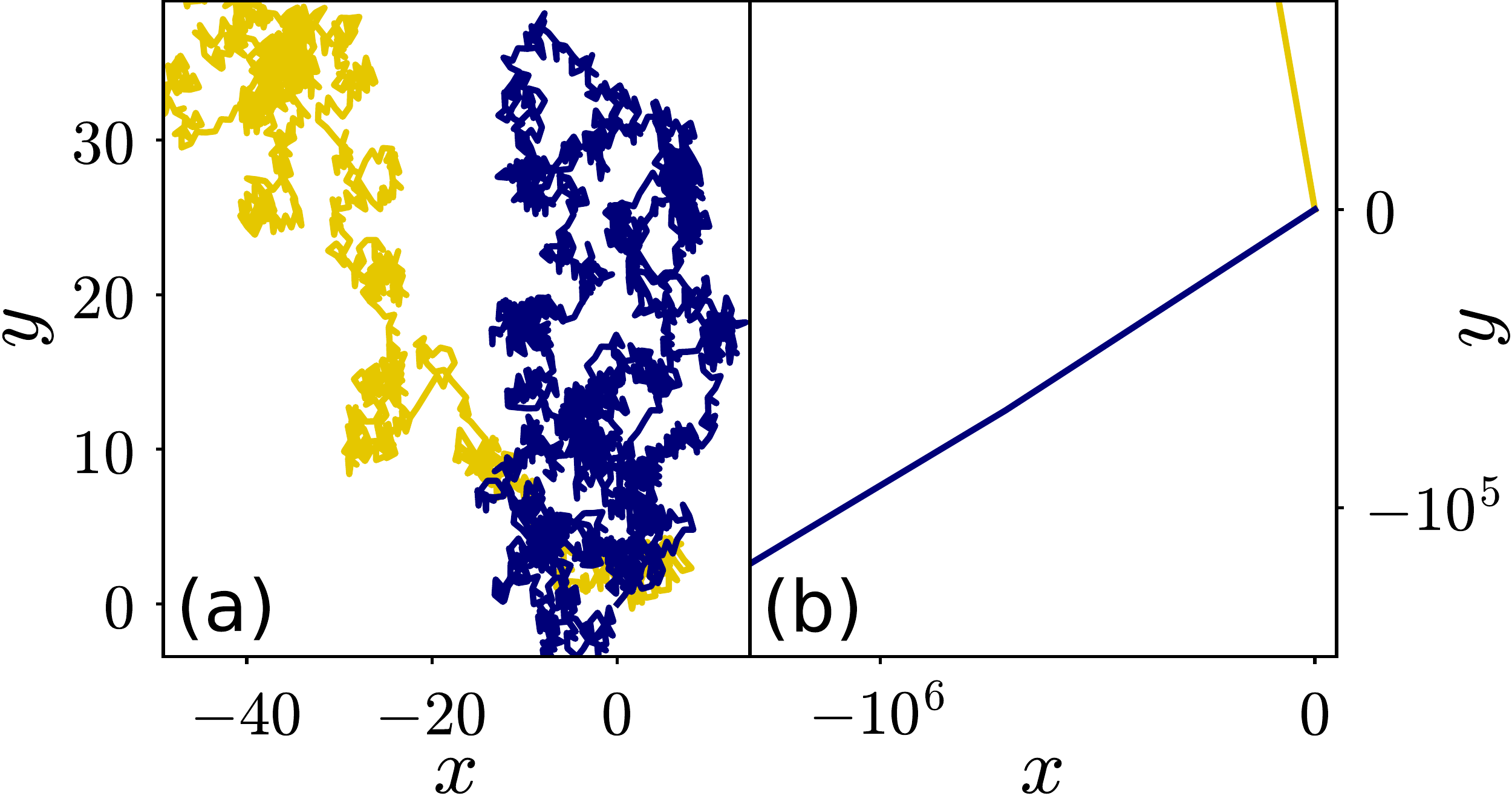}}
\caption{
Trajectory of chaotic motion for a tendency of alignment $A=9.940441$ \am{for an angular frequency $\omega=\pi/5$ (darker lines)}. %In this case, 
The order parameter in Eq.~(\ref{eq_pd}) assumes a very low value of $p\approx0.0066$, see also the square in Fig.~\ref{fig_arnold}(b). 
%6.56\times10^{-3}$. 
(a) An irregular shape appears on shorter time and length scales. %, here for the first $...$ time steps. 
(b) Conversely, the net motion towards one direction is visible on large time and length scales, %here for the first $...$ time steps, 
where the trajectory appears rather straight. %again.
\am{For comparison, evaluations for $\omega=\pi/\sqrt{26}$ are included as well (brighter lines). \rev{Although the curves appear qualitatively similar in shape, the small change in $\omega$ leads for the chosen magnitude of $A$ to significant quantitative deviations, particularly concerning the overall drift angle in (b).}}
}
\label{fig_exampledrift}
\end{figure}

To demonstrate this aspect, we concentrate on %the parameter value 
$A=9.940441$, for which $p\approx0.0066$ is very low. \am{Here, significant amounts of overreaction occur in individual steps of angular alignment already for weak deviations from the heading towards $-\mathbf{\hat{x}}$}. We plot the corresponding trajectory in Fig.~\ref{fig_exampledrift}(a) \am{on a shorter time scale. Previous studies focusing on variants of Eq.~(\ref{eq_varphi}) have reported diffusive behavior of the angular variable $\varphi$ for $\omega=0$ \cite{geisel1982onset, schell1982diffusive}. Considering here in addition the spatial variables,} a net drift becomes obvious when we increase the time interval. Zooming out, the trajectory more and more resembles again a straight line, see Fig.~\ref{fig_exampledrift}(b). 
%
%Since nevertheless the trajectories are still deterministic, ensemble averages are not really meaningful. Instead, we average over initializations by different angles $\varphi_0$. %, not only $\varphi_0=0$. 
%\am{Thus, at least for the investigated parameter values of $A$, we }
Averaging as a test over $1000$ different initializations of $\varphi_0$ in the interval of $[0,2\pi[$, we still obtain $\langle p\rangle\approx0.0086$, which agrees with the order of magnitude mentioned above. 

Eventually, the order parameter $p$ does not quantify whether the motion is headed towards the requested direction $-\mathbf{\hat{x}}$. For this reason, we introduce a second order parameter
\begin{equation}\label{eq_prd}
p_{\am{-\mathbf{\hat{x}}}}=\frac{1}{2N_{\mathrm{step}}}\sum_{n=1}^{N_{\mathrm{step}}}\left(1-\cos\varphi_n\right).
\end{equation}
We obtain $p_{\am{-\mathbf{\hat{x}}}}=1$ when the motion is fully directed towards $-\mathbf{\hat{x}}$ during each time step, while $p_{\am{-\mathbf{\hat{x}}}}$ vanishes for straight motion into the opposite direction $+\mathbf{\hat{x}}$. 
\am{In between, $p_{\am{-\mathbf{\hat{x}}}}=0.5$ signals that there is no net motion along $\mathbf{\hat{x}}$.} $p_{\am{-\mathbf{\hat{x}}}}$ is indicated in Fig.~\ref{fig_arnold}(b) by the brighter line. 
Apparently, completely directed motion along a straight line does not necessarily imply an effective motion towards the target direction $-\mathbf{\hat{x}}$. \rev{In our situation, we address actively driven objects of given persistent bends in their trajectories and given strengths of alignment tendency as quantified by a set magnitude of $\omega$ and $A$, respectively. We thus conclude that, to achieve for such prescribed values a most effective requested propagation towards a target, it may be reasonable to work with an angular offset for the heading direction.} %That is, a heading towards an oblique direction is programmed to reach most effectively a target at $-\mathbf{\hat{x}}$. 

\section{Collective dynamics}
\label{sec_coll}

Finally, we investigate how mutual interactions %between self-propelled or actively driven objects 
support collective and rectified motion, particularly in the chaotic regime. To this end, we consider orientational interactions of the Vicsek type \cite{vicsek1995novel} between $N$ self-propelled or actively driven 
objects, labeled by $i=1,...,N$. %Before evaluating for each object separately Eq.~(\ref{eq_varphi}), 
%That is, we introduce %for each object $i$ a 
%steps of reorientation. % affected by all nearby objects within a distance $d$. 
At each time step, the angle $\varphi_i$ of the $i$th object is set equal to the averaged heading of all objects within a distance $d$ from it. 
\am{Obvious example systems concerning discrete collective alignment interactions are robotic swarms or otherwise artificial, programmed realizations of actively driven objects that suffer from finite refresh rates of their sensors \cite{vasarhelyi2018optimized}. If monitored or steered through a common, central unit \cite{bauerle2018self, lavergne2019group}, the update may become rather simultaneous, or at least be based on simultaneously gathered information.}

To obtain our results, \am{we begin by considering only the deterministic contributions described so far. We} evaluate the mentioned Vicsek-type \am{alignment} interactions before %evaluating 
Eq.~(\ref{eq_varphi}). Yet, we have checked that our results are qualitatively identical when switching this order.  
As an initialization, we iterate $\varphi_i$ for the $i$th object $iN$ times according to Eq.~(\ref{eq_varphi}). Only then mutual angular interactions and transport %as described by Eqs.~(\ref{eq_x}) and (\ref{eq_y}) 
are introduced. The $N$ objects are initially distributed at random. They are confined to a square-like box of side length $L$ under periodic boundary conditions. 
We here set $\omega=\pi/5$, $A=9.940441$, $d=1$ identical for all objects, and $L=10$. Increasing the number of objects $N$ up to $1000$, we have not identified qualitative variations in the collective behavior. 

For quantification, we denote by $\varphi_{i,n}$ the angular orientation of the $i$th object at the $n$th time step and evaluate the order parameters 
\begin{equation}\label{eq_Pde}
P_{\am{\mathrm{step}}}=\frac{1}{NN_{\mathrm{step}}}\!\!\sum_{n=1}^{N_{\mathrm{step}}}\!\left[\!\left(\sum_{i=1}^{N}\cos\varphi_{i,n}\!\right)^{\!\!\!2}\!\! +\! \left(\sum_{i=1}^{N}\sin\varphi_{i,n}\!\right)^{\!\!\!2}\, \right]^{\!1/2}\!\!\!,
\end{equation}
\begin{eqnarray}
P&=&\frac{1}{NN_{\mathrm{step}}}\left[\left(\sum_{n=1}^{N_{\mathrm{step}}}\sum_{i=1}^{N}\cos\varphi_{i,n}\!\right)^{\!\!\!2}\! \right.
\nonumber\\[-.2cm]
&&{}
\hspace{2.5cm}+\! \left.\left(\sum_{n=1}^{N_{\mathrm{step}}}\sum_{i=1}^{N}\sin\varphi_{i,n}\!\right)^{\!\!\!2}\: \right]^{\!1/2}\!\!\!,
\hspace{.3cm}
\label{eq_Pd}
\end{eqnarray}
and
\begin{equation}\label{eq_Prd}
P_{\am{-\mathbf{\hat{x}}}}=\frac{1}{2\,NN_{\mathrm{step}}}\sum_{n=1}^{N_{\mathrm{step}}}\sum_{i=1}^{N}\left(1-\cos\varphi_{i,n}\right).
\end{equation}
%as depicted in Fig.~\ref{fig_collective}. 
$P_{\am{\mathrm{step}}}$ basically measures whether (solely) at each time step all objects move into the same direction, no matter whether this direction changes over the considered overall period. % first determines at each time step the degree of coordinated collective motion from all $N$ objects. This scalar degree of order is then averaged over the considered time steps. 
Conversely, $P$ determines the degree of directed collective motion not distinguishing between different objects and %different 
time steps. It decreases when over time the direction of ordered collective motion changes. 
$P_{\am{-\mathbf{\hat{x}}}}$ again quantifies the effectiveness of motion along the preferred direction $-\mathbf{\hat{x}}$. %, averaged over all $N$ objects and $N_{\mathrm{step}}$ time steps.

\begin{figure}
\centerline{\includegraphics[width=.9\columnwidth]{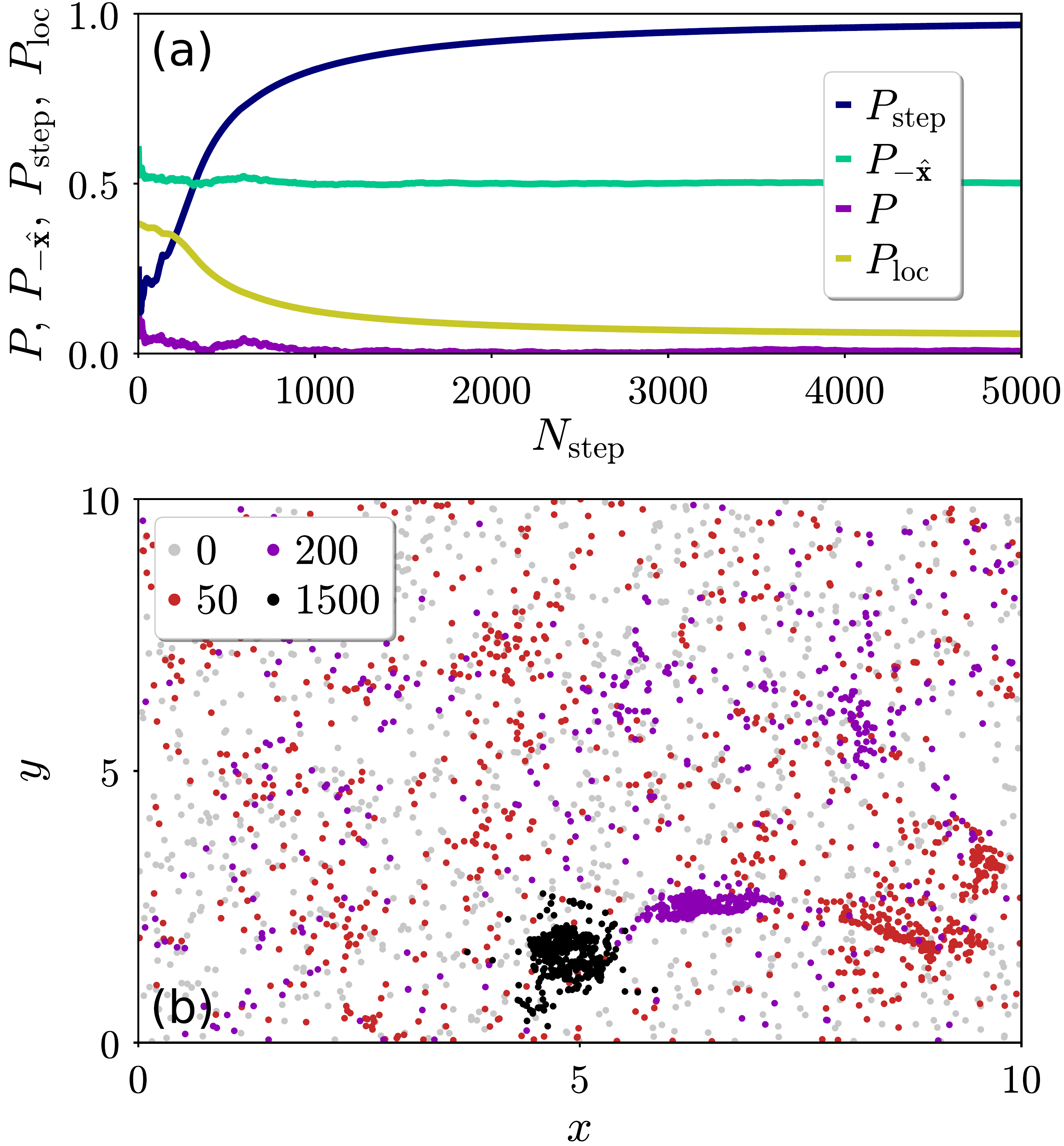}}
\caption{
Emergence of ordered collective motion in a crowd of $N=1000$ initially randomly arranged objects in the chaotic regime for $\omega=\pi/5$ and $A=9.940441$, see also Fig.~\ref{fig_exampledrift}. Vicsek-type mutual interactions of range $d=1$ in a periodic box of length $L=10$ are included. (a) Time evolution of the order parameters $P_{\am{\mathrm{step}}}$, $P$, and $P_{\am{-\mathbf{\hat{x}}}}$, see Eqs.~(\ref{eq_Pde})--(\ref{eq_Prd}), indicates development of orientationally ordered collective motion ($P_{\am{\mathrm{step}}}$). Yet, the overall direction of collective motion is not correlated over time ($P$) and not effectively oriented into the desired direction ($P_{\am{-\mathbf{\hat{x}}}}$). Here, $N_{\mathrm{step}}$ always corresponds to the total number of simulated time steps. \rev{The development of spatial concentration of the objects is quantified by the order parameter $P_{\mathrm{loc}}$, see Eq.~(\ref{eq_Ploc}).} (b) Positions of the objects after the amount of time steps indicated in the legend. A concentration from initial random distribution into a spot of approximate size $d=1$ is obvious. 
\am{This spot basically moves as one entity during the further course of the dynamics \cite{suppl}.}
}
\label{fig_collective}
\end{figure}

We generally observe that the initial diversification in angular distribution does not survive the averaging procedure \am{when stochastic contributions are absent}. In fact, the order parameter $P_{\am{\mathrm{step}}}$ approaches values close to one after an initial period of ordering in all considered cases, see Fig.~\ref{fig_collective}(a). That is, at each time step (separately), all objects move basically into the same direction. 
However, here for $A=9.940441$, this direction becomes uncorrelated in time, and %the order parameter 
$P$ drops towards zero. It appears as if the whole crowd synchronizes and moves chaotically as one entity. %In agreement with the selected value of $A$, this motion is rather uncorrelated in time, as indicated by $P$. 
Moreover, the overall motion is hardly directed into the direction $-\mathbf{\hat{x}}$, as signaled by $P_{\am{-\mathbf{\hat{x}}}}\,\am{ \approx 0.5}$, see Fig.~\ref{fig_collective}(a). 

Interestingly, orientational ordering for $A=9.940441$ is associated with spatial concentration of the objects. % within an extension of order $d$. 
We found this process of concentration independently of the number of objects $N$, and in all cases we started from a random spatial distribution throughout the periodic box. Thus, the spatial and orientational dynamics are significantly coupled. An example of initial distribution and subsequent spatial concentration in one spot is depicted in Fig.~\ref{fig_collective}(b). 
\am{%Indeed, the motion of the objects in the concentrated spot is highly correlated. 
This spot basically moves as one entity during further iteration, as illustrated in a video in the Supplemental Material \cite{suppl}.} 
\rev{To quantify this localization into one spot, we introduce another order parameter, 
\begin{eqnarray}
P_{\mathrm{loc}} &=& 
\frac{1}{N^2N_{\mathrm{step}}L}\sum_{n=1}^{N_{\mathrm{step}}}\sum_{i=1}^N\sum_{j=1}^N \Big[ (x_{i,n}-x_{j,n})^2
\nonumber\\
&&{}\qquad\qquad\qquad  + (y_{i,n}-y_{j,n})^2 \Big]^{1/2}.
\label{eq_Ploc}
\end{eqnarray}
For perfect concentration of all objects into one dot, we would find $P_{\mathrm{loc}}=0$. Otherwise, $P_{\mathrm{loc}}$ is the larger in magnitude the more spatially distributed the particles are. During evaluation, we take into account the periodic boundary conditions and always use the minimal distance between any two objects, which, depending on the situation, may be measured across the periodic boundaries.}

\am{So far, we have regarded the Vicsek-type alignment interactions as completely deterministic. In reality and generally, fluctuations and errors occur. Similarly to the Vicsek model \cite{vicsek1995novel}, we therefore include a stochastic contribution $\Gamma_{i,n}$ that is added to the angle $\varphi_{i,n}$ after evaluating the alignment interactions but before evaluating Eq.~(\ref{eq_varphi}). It accounts, for example, for the imperfections arising during the alignment procedure. We assume $\Gamma_{i,n}$ to be uncorrelated in time, white, Gaussian distributed, and uncorrelated between the objects, so that $\langle\Gamma_{i,n}\rangle=0$ and $\langle\Gamma_{i,n}\,\Gamma_{j,m}\rangle=2K\delta_{ij}\delta_{nm}$. Here, $\delta$ denotes the Kronecker delta and $K$ sets the strength of the stochastic contribution.}

\begin{figure}
\centerline{\includegraphics[width=.9\columnwidth]{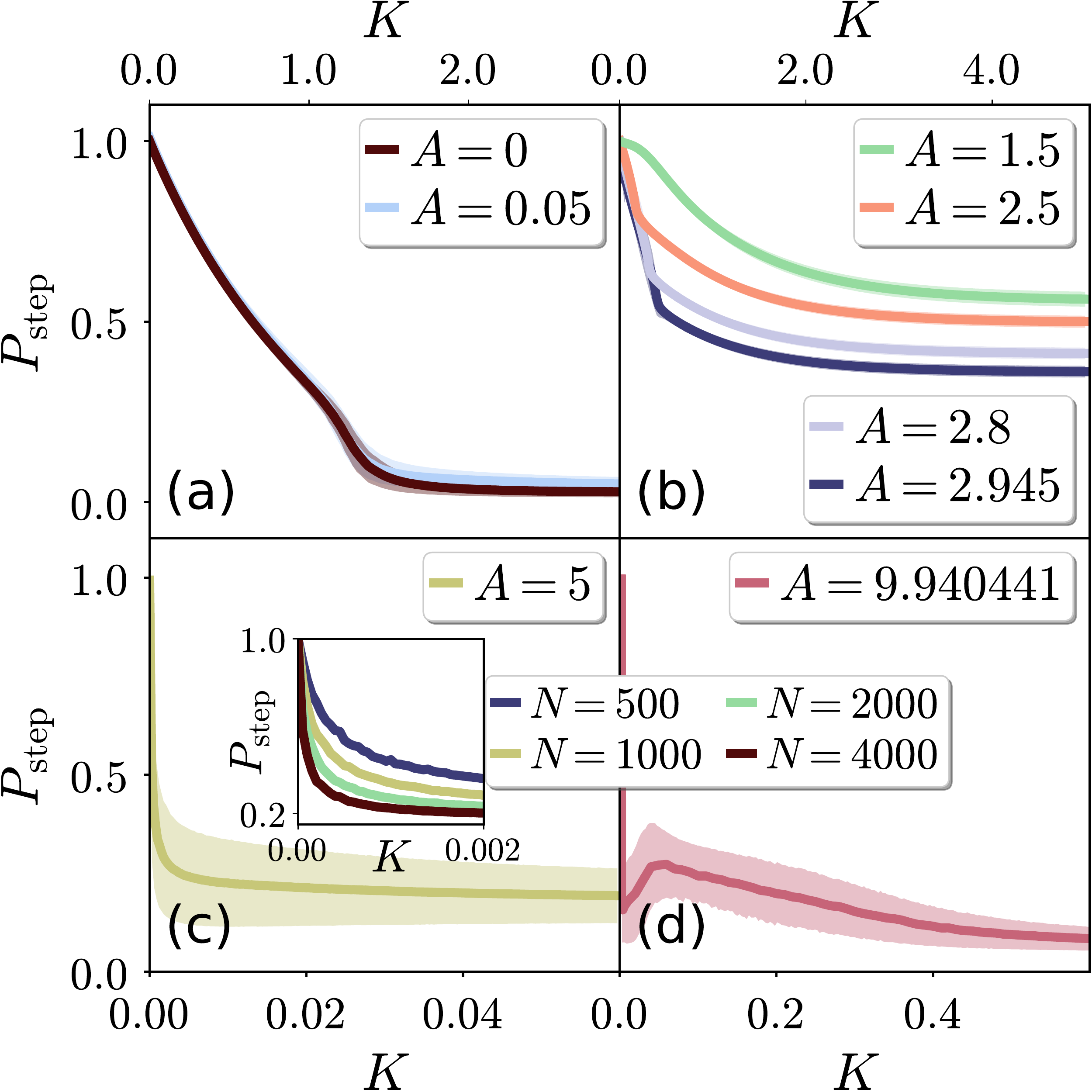}}
\caption{
\am{
Influence of additional stochastic contributions of strength $K$ to the angular alignment of each object, superimposed before evaluating Eq.~(\ref{eq_varphi}), on the magnitude of the resulting order parameter $P_{\mathrm{step}}$ in a crowd of $N=1000$ objects and $\omega=\pi/5$. For illustration, the values of $A$ are set to (a) $0$ and $0.05$, (b) $1.5$, $2.5$, $2.8$, and $2.945$, (c) $5$, and (d) $9.940441$, as in Figs.~\ref{fig_simpletraj}(a), \ref{fig_complextraj}, and \ref{fig_exampledrift}. Apparently, the sensitivity with respect to stochastic contributions significantly increases in the chaotic regime. \rev{Standard deviations in the main plots are indicated by light shaded areas in the background and were obtained by sampling over $N_{\mathrm{step}}=5\times10^4$ iterative steps. % in (a) and (b) while to this end $N_{\mathrm{step}}=2.5\times10^5$ in (c) and (d). 
The inset in (c) magnifies the drop of the order parameter $P_{\mathrm{step}}$ with increasing low magnitude of $K$ at constant density but varying total number of objects $N$.}
}
}
\label{fig_K}
\end{figure}

We numerically measured the time evolution of the order parameters choosing, for illustration, for $\omega=\pi/5$ the same values of $A$ as in Figs.~\ref{fig_simpletraj}(a), \ref{fig_complextraj}, and \ref{fig_exampledrift}. Corresponding results for the magnitude of the order parameter $P_{\mathrm{step}}$, see Eq.~(\ref{eq_Pde}), as a function of the stochastic strength $K$ are depicted in Fig.~\ref{fig_K} for one stochastic realization for each value of $K$, yet after averaging over at least the last $N_{\mathrm{step}}=5\times10^4$ time steps of each simulated realization.%

For vanishing and low tendency of alignment $A$, see Fig.~\ref{fig_K}(a), we observe what might be expected from the Vicsek model. The whole crowd synchronizes for smaller stochastic strengths $K$. With growing values of $K$, this ordering breaks down. %The quick disordering with increasing $K$ in Fig.~\ref{fig_K}(a) can be ascribed to the pronounced circling behavior for low tendency of external alignment $A\approx0$.

\am{
Raising the tendency of alignment $A$ into the regime of the first bifurcations in Fig.~\ref{fig_arnold}(a), see Fig.~\ref{fig_K}(b), the sensitivity with respect to the stochastic contributions increases \rev{when we increase $A$ from $1.5$ to $2.945$}. In principle, the stochastic influence on an object can push its angular orientation to another bifurcated branch. Overall, ordering is then affected.} \rev{For the depicted values of $K$, the order parameter $P_{\mathrm{step}}$ remains at finite, nonvanishing magnitude, because in our consideration the stochastic contributions $\Gamma_{i,n}$ are predominantly considered to arise from the mutual Vicsek-type alignment interactions between the objects. The stochastic terms are thus included before evaluating Eq.~(\ref{eq_varphi}), as mentioned above. Therefore, the actual individual tendency of alignment due to $A\neq0$ in Eq.~(\ref{eq_varphi}) remains and tends to drive all objects individually towards the preferred direction. As a result, nonvanishing overall orientational order $P_{\mathrm{step}}$ arises, despite the stochastic fluctuations in mutual alignment between the objects. Switching this order and adding the stochastic contributions only after evaluating Eq.~(\ref{eq_varphi}), the curves, for instance, in Fig.~\ref{fig_K}(b) drop further towards zero with increasing $K$.}

The trend of decaying order parameter $P_{\mathrm{step}}$ naturally becomes still more pronounced for tendencies of alignment $A$ in the chaotic regime, see Figs.~\ref{fig_K}(c) and (d). Even if only weak deviations in the angular orientation of an individual object are caused by the stochastic contributions, they here according to Eq.~(\ref{eq_varphi}) can result in significant angular deviations during subsequent time steps. A bias of this type strongly counteracts overall synchronization. Consequently, very weak stochastic influences can already induce a breakdown in overall alignment. 
\rev{The inset in Fig.~\ref{fig_K}(c) demonstrates that the drop in the order parameter $P_{\mathrm{step}}$ is still continuous at the considered system sizes at small magnitudes of the strength of stochastic contributions $K$. Yet, with increasing system size, that is, for increasing total number of objects $N$ at identical overall density, the drop becomes steeper. At this point, we cannot exclude that an actual transition occurs from $K=0$ to $K>0$ for infinitely extended systems.}

\begin{figure}
\centerline{\includegraphics[width=.9\columnwidth]{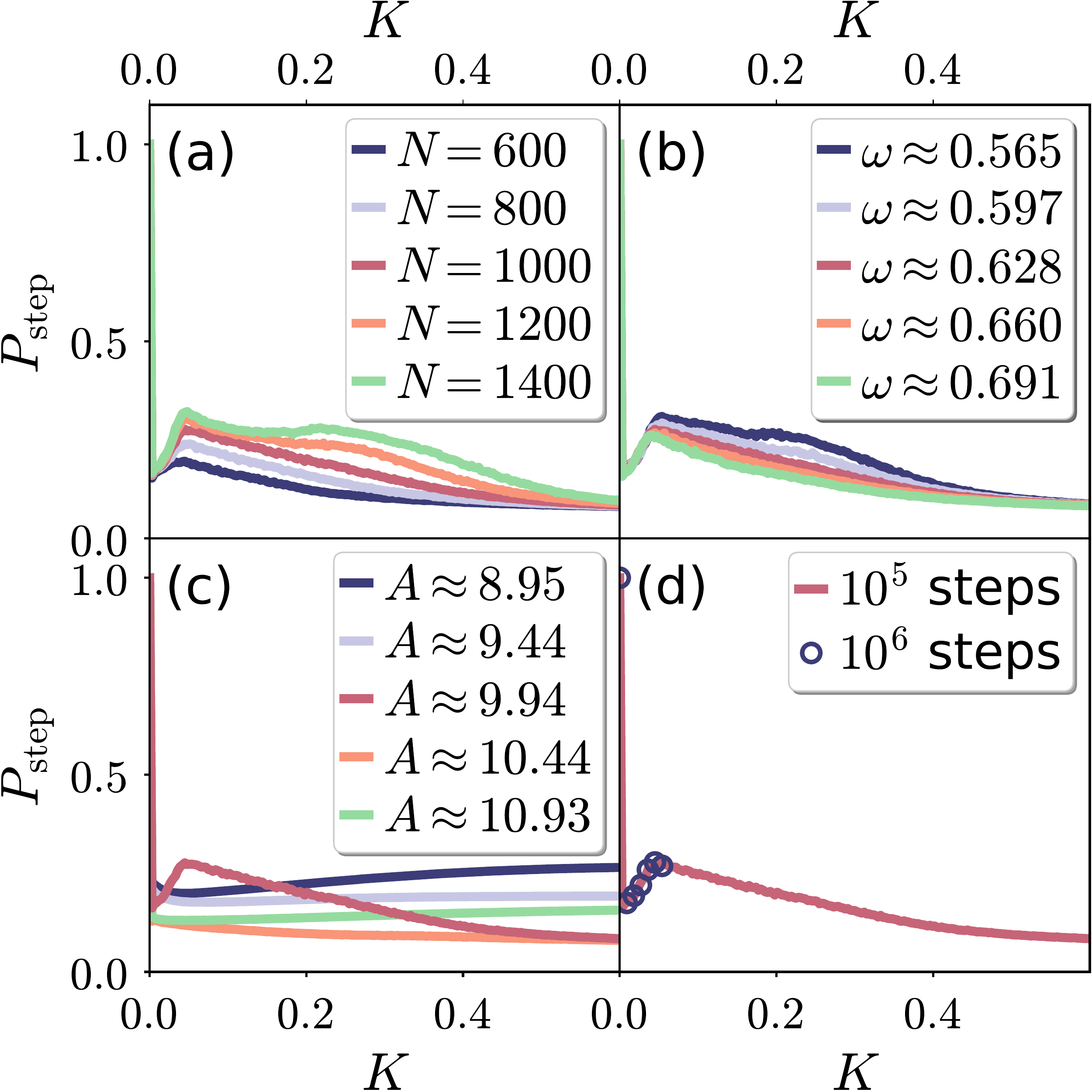}}
\caption{
\rev{
Stability of the results in Fig.~\ref{fig_K}(d) against variations (a) in the number of objects $N$, (b) in the angular frequency $\omega$, (c) in the strength of alignment tendency $A$, and (d) in the number of steps of iteration before the order parameter $P_{\mathrm{step}}$ was recorded. Here, we set $N_{\mathrm{step}}=5\times10^4$, and in (d) the legend displays the overall number of iterative steps in that panel. Standard deviations are omitted for better visibility. 
}
}
\label{fig_nonmono}
\end{figure}

Interestingly, we observe a nonmonotonic curve in Fig.~\ref{fig_K}(d). After a steep decrease of the order parameter at very small amplitudes of fluctuations, it slightly recovers when we continue to increase the magnitude of fluctuations. Accordingly, in this chaotic example with significant overreaction, the presence of weak fluctuations can actually support ordering. Afterwards, with further increasing magnitude of fluctuations, the order parameter continues to decrease.
\rev{Our analysis revealed that this nonmonotonic behavior is stable against moderate variations in the number of objects $N$ and in the angular frequency $\omega$, see Figs.~\ref{fig_nonmono}(a) and (b), respectively. However, the effect is significantly more sensitive with respect to changes in the strength of alignment tendency $A$, see Fig.~\ref{fig_nonmono}(c). In Fig.~\ref{fig_nonmono}(d), we confirm that the phenomenon is not associated with an insufficient amount of steps of iteration before evaluating the order parameter.}%

\begin{figure}
\centerline{\includegraphics[width=.9\columnwidth]{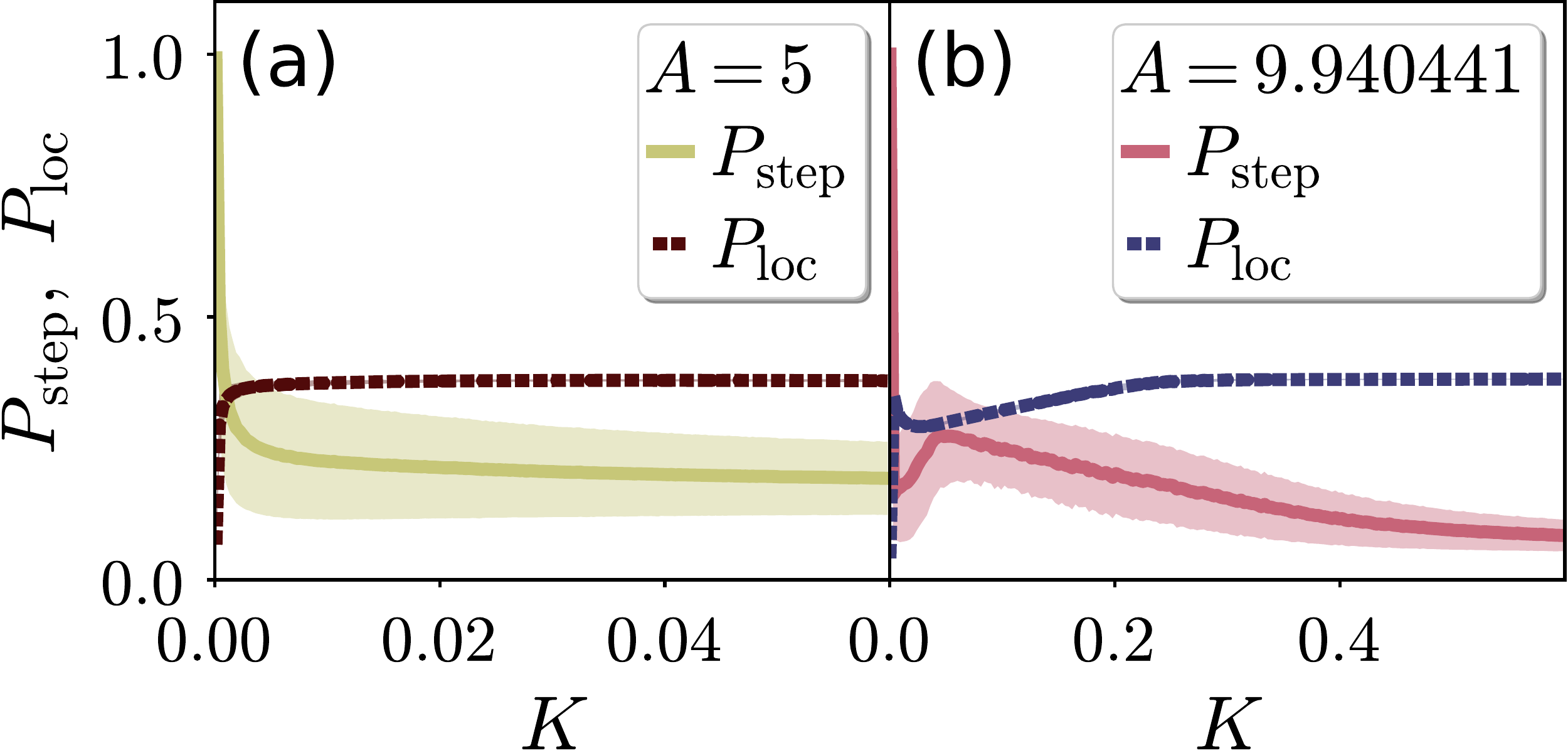}}
\caption{
\rev{
Same as Figs.~\ref{fig_K}(c) and (d), but now amended by the magnitude of spatial concentration quantified by the order parameter $P_{\mathrm{loc}}$, see Eq.~(\ref{eq_Ploc}). As in Figs.~\ref{fig_K}(c) and (d), the values of $A$ are set to (a) $5$ and (b) $9.940441$, respectively. Standard deviations were obtained in the same way as indicated in the caption of Fig.~\ref{fig_K} and here for $P_{\mathrm{loc}}$ at maximum are of the order of the line thickness. 
}
}
\label{fig_Ploc}
\end{figure}

\am{
Briefly, we address the consequences of stochastic fluctuations during mutual alignment on the spatial concentration process indicated in Fig.~\ref{fig_collective}(b). In that case, we observe a reduced trend of spatial concentration into one spot and of its correlated motion, as illustrated in the videos in the Supplemental Material \cite{suppl}. \rev{Quantitatively, we have measured these trends using the order parameter $P_{\mathrm{loc}}$ as defined in Eq.~(\ref{eq_Ploc}). The results are illustrated in Fig.~\ref{fig_Ploc} in the chaotic regime for $A=5$ and $A=9.940441$. Here, the increase (decrease) in $P_{\mathrm{loc}}$ is approximately correlated with the decrease (increase) in $P_{\mathrm{step}}$. At $K=0$, the low magnitude of $P_{\mathrm{loc}}$ indicates significant spatial concentration, which we have not observed for the other values of $A$ studied in Fig.~\ref{fig_K}. At elevated values of $K$, the order parameter $P_{\mathrm{loc}}$ approaches the magnitude expected for approximately equally spatially distributed objects in our system.} An interesting question for future investigations concerns the relation between this spatial concentration and the propagating structures of high density emerging in regular Vicsek models \cite{chate2008collective, chate2008modeling, menzel2012collective, ihle2013invasion}.}

\am{
Finally, we consider the polydispersity of the actively driven objects as a source of disorder, instead of fluctuations in mutual alignment interactions. To this end, we concentrate on the tendency of alignment along $-\mathbf{\hat{x}}$, parameterized by $A$. Specifically, before the angular and spatial initialization, we assign to each object a value $A_i=A+\Gamma_{A,i}$ that is kept constant during the numerical iteration ($i=1,...,N$). $\Gamma_{A,i}$ is drawn from a Gaussian distribution and uncorrelated between different objects, implying $\langle\Gamma_{A,i}\rangle=0$ and $\langle\Gamma_{A,i}\,\Gamma_{A,j}\rangle=2\,K_A\,\delta_{ij}$.
}

\begin{figure}
\centerline{\includegraphics[width=.9\columnwidth]{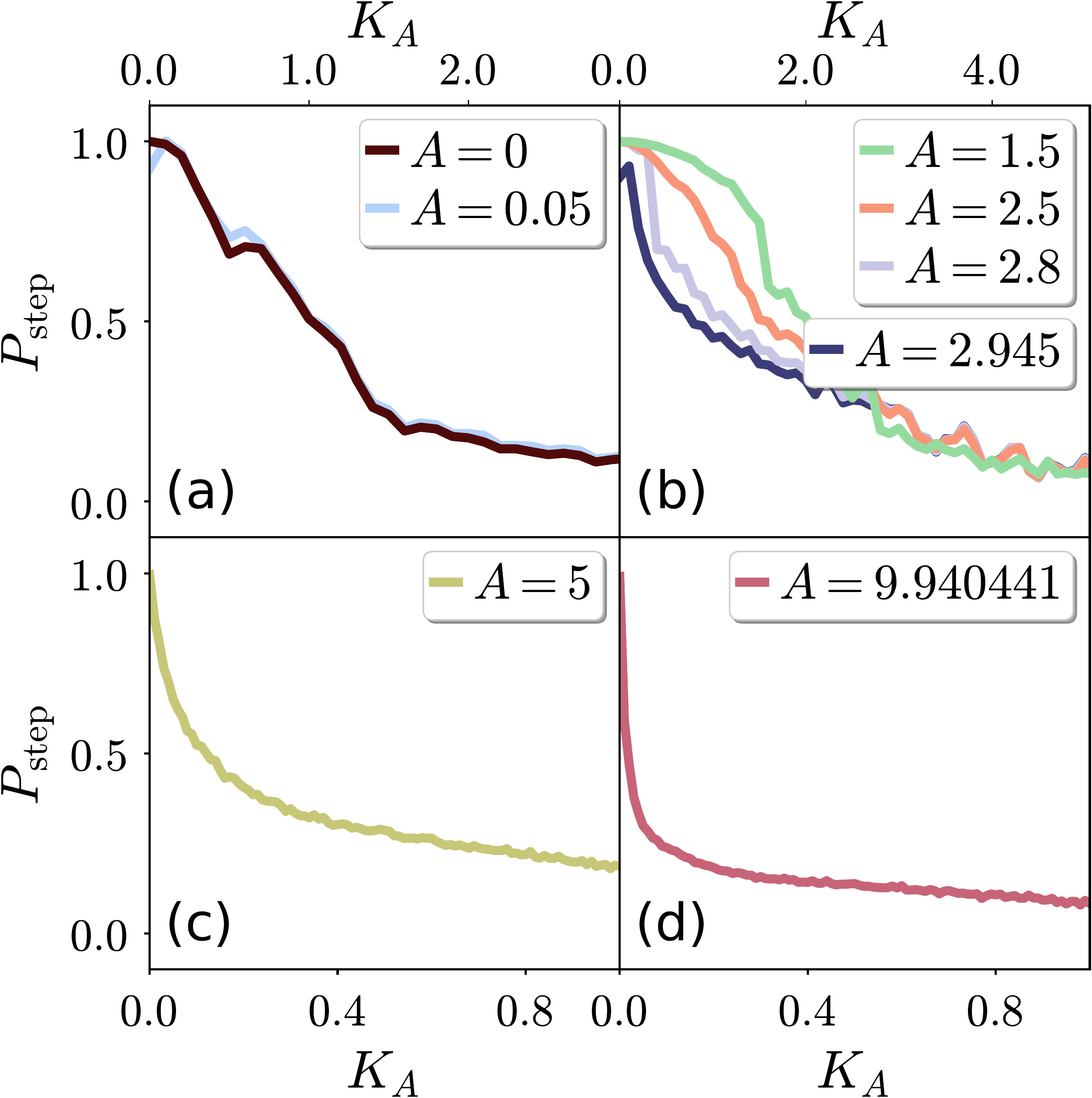}}
\caption{
\am{
Influence of the magnitude $K_A$ of polydispersity in the tendency of alignment $A$ on the resulting order parameter $P_{\mathrm{step}}$ in a crowd of $N=1000$ objects and $\omega=\pi/5$. One specific realization of polydispersity is addressed for each value of $K_A$. Again, for comparison, the values of $A$ are set to (a) $0$ and $0.05$, (b) $1.5$, $2.5$, $2.8$, and $2.945$, (c) $5$, and (d) $9.940441$, as in Figs.~\ref{fig_simpletraj}(a), \ref{fig_complextraj}, \ref{fig_exampledrift}, and \ref{fig_K}--\ref{fig_Ploc}. The sensitivity with respect to polydispersity apparently is more pronounced in the chaotic regime. \rev{Standard deviations obtained by averaging over five different realizations of the systems are mostly of the order of the line thickness or smaller.}
}
}
\label{fig_A}
\end{figure}

\am{
To evaluate the influence of polydispersity on our results, we increase $K_A$ for the same values of $A$ studied before in Figs.~\ref{fig_simpletraj}(a), \ref{fig_complextraj}, \ref{fig_exampledrift}, and \ref{fig_K}--\ref{fig_Ploc}. Plotting the order parameter $P_{\mathrm{step}}$ as a function of increasing magnitude of polydispersity $K_A$, here for one different realization for each value of $K_A$, we infer from Fig.~\ref{fig_A} similar qualitative trends as for increasing strengths of fluctuations in mutual angular alignment in Fig.~\ref{fig_K}, except for the nonmonotonic behavior in Fig.~\ref{fig_K}(d).} \rev{However, quantitatively, the decrease in Fig.~\ref{fig_A}(b) is more substantially towards zero than in Fig.~\ref{fig_K}(b). Specifically when varying the tendency of alignment $A$ for the individual objects in Fig.~\ref{fig_A}, different preferred individual angles of alignment result, see Fig.~\ref{fig_arnold}. In effect, this naturally reduces their overall orientational ordering as well, and $P_{\mathrm{step}}$ is reduced.} Our example systems become particularly sensitive concerning polydispersity in the chaotic regime. 

\section{Conclusions}
\label{sec_concl}

Summarizing, we have investigated the motion of self-propelled particles or actively driven objects that, when undisturbed, move in discrete steps along regular kinked trajectories. Rich nonlinear dynamics emerges when an additional tendency of alignment along a specified direction comes into play. %In reality, this might be a remote source of nutrient that attracts self-propelled organisms that perform stepwise migration. 
%The parameter that sets the strength of alignment turns out to play a crucial role. 
We have observed discrete versions of cycloidal-like trajectories, % that resemble closed polygon-like trajectories with an additional drift, as expected. However, additional types of motion appear. Examples are 
straight motion under an oblique drift angle, various zigzag-like modes of migration, as well as chaotic motion in broad parameter intervals.  %. Remarkably, in broad parameter intervals, the motion becomes chaotic. 

In fact, we found that this generic type of migration is associated with one of the most basic minimal models of nonlinear dynamics, namely, the Arnol'd circle map \cite{arnold2009circle}. Through active transport, this nonlinear dynamics gets laid out into the two-dimensional plane in the form of the resulting trajectories. We always observe a net drift along a certain direction, even if comparatively weak and not necessarily along the requested orientation. Consequently, to reach a target at a certain remote location in combination with \rev{a set} tendency towards circular trajectories, it generally becomes reasonable to work with an angular offset. Besides, the tendency of alignment must be tuned carefully. 

This simple example demonstrates the richness that nonequilibrium and actively driven systems feature already in a very basic context. %We hope to stimulate by our considerations corresponding experimental investigations. 
%A biological example may be given by imperfect alga cells of \textit{Chlamydomonas reinhardtii} when showing pronounced differences in the performance of their two flagella \cite{brokaw1982analysis, kamiya1984submicromolar} and triggered to propel towards a certain direction \cite{schaller1997chlamydomonas}. They swim by discrete strokes. 
\am{To observe the described effects in reality,} nonsymmetric chiral hoppers driven through vibrating substrates can be biased towards one direction by gravity when inclining the vibrating plate \cite{goohpattader2010diffusive, mohammadi2020dynamics}.

\am{As a final remark,} we mention that the picture of a drunk person trying to get home or to the next pub is frequently used to motivate the concept of a random walk. \am{One may argue about whether this comparison is generally reasonable. After all, the person has a remote target and tries to reach it. Therefore, in} view of our results, it may be \am{more appropriate} to reinterpret this type of motion of a person having the spins, performing discrete steps, and trying to reach a target %. Whether intoxicated or not, the movement may rather need to be regarded 
as chaotic rather than random.

\begin{acknowledgments}
%\vspace{-.4cm}
The author thanks the Deutsche Forschungsgemeinschaft (German Research Foundation, DFG) for support through the Heisenberg Grant No.~ME 3571/4-1. 
\end{acknowledgments}
%\vspace{-.3cm}

%\section*{Data availability}
%%%\vspace{-.4cm}
%The data that support the findings of this study are available within the article and/or result from solving Eqs.~(\ref{eq_Lang_v})--(\ref{eq_fp}) as described in the text.
%%%\vspace{-.3cm}

%\section*{Author declaration}
%%%\vspace{-.4cm}
%The author has no conflicts to disclose. 
%%%\vspace{-.3cm}

%\section*{References}
%%%\appendix

%\vspace{-.7cm}

%\bibliography{references}


\end{document}